\documentclass[12pt,preprint]{aastex}
\usepackage{amsmath}
\newcommand{\pd}{\partial}
\newcommand{\lowthres}{0.01\%}
\newcommand{\medthres}{2\%}
\newcommand{\highthres}{60\%}

\shorttitle{Solar Particle Injection Timing}
\shortauthors{S{\' a}iz et al.}

\slugcomment{Astrophys. J., in press}

\begin{document}

\title{On the Estimation of Solar
    Energetic Particle Injection Timing from Onset Times near Earth}

\author{
Alejandro S\'aiz,\altaffilmark{1,2}
Paul Evenson,\altaffilmark{3}
David Ruffolo,\altaffilmark{1} and
John W. Bieber\altaffilmark{3}
}
 \altaffiltext{1}{Department of Physics, Faculty of Science, Mahidol University, Bangkok 10400, Thailand}
 \altaffiltext{2}{Department of Physics, Faculty of Science, Chulalongkorn University, Bangkok 10330, Thailand}
 \altaffiltext{3}{Bartol Research Institute, University of Delaware, Newark, DE 19716}

\begin{abstract}

We examine the accuracy of a common technique for estimating the
start time of solar energetic particle injection based on a linear
fit to the observed onset time versus (particle velocity)$^{-1}$.
This is based on a concept that the first arriving particles move
directly along the magnetic field with no scattering.  We check
this by performing numerical simulations of the transport of solar
protons between~2 and 2000~MeV from the Sun to the Earth, for
several assumptions regarding interplanetary scattering and the
duration of particle injection, and analyzing the results using
the inverse velocity fit.  We find that in most cases, the onset
times align close to a straight line as a function of inverse
velocity.  Despite this, the estimated injection time can be
in error by several minutes.  Also, the estimated path length can
deviate greatly from the actual path length along the
interplanetary magnetic field. The major difference between the
estimated and actual path lengths implies that the first arriving
particles cannot be viewed as moving directly along the
interplanetary magnetic field.
\end{abstract}

\keywords{Sun: particle emission --- interplanetary medium
--- solar-terrestrial relations --- methods: numerical}

\section{Introduction}

An important issue when studying solar events is the exact time
when energetic particles are first released from the Sun or its
vicinity. This is crucial to understanding the mechanisms of
particle acceleration and where it takes place
\citep[e.g.,][]{Kahleretal90}. When inferring the start time of
particle release, $t_0$, one has to take into account the many
different processes acting on the particles from their release
until the time of detection, $t_{\rm onset}$, at spacecraft or
Earth-based instruments. First the particles (most of them
protons) are released with a finite duration of injection at the
Sun, and with velocity $v=\beta c$. Because they are charged
particles, their motion mostly follows the interplanetary magnetic
field, gyrating with a pitch angle $\theta$ (defined as the angle
between the velocity and the mean field). At the same time, the
particles suffer interplanetary scattering due to resonant
interactions with magnetic field irregularities leading to a
random walk in pitch angle \citep{Jokipii66}, thus varying the
component of the velocity in the direction of the mean magnetic
field, $v_z=v\cos\theta$. The random walk can even lead to
$\theta>90^\circ$, which means that particles move back toward the
Sun. This leads to a random walk in position as well, and hence to
spatial diffusion. There are also effects due to the solar wind
speed, such as convection and adiabatic deceleration. These
various effects can be taken into account to precisely determine
the start time of injection at the Sun
\citep[e.g.,][]{Bieberetal02,Bieberetal04,Oct28}.

A popular approximation is to consider that the first observed
particles move approximately parallel to the mean magnetic field
(with $v_z=v$). By doing this one neglects the effects of
interplanetary scattering at onset. Then an apparently easy way to
estimate the timing is to observe the time of detection onset and
shift it according to the path length travelled by the particles.
This path length is typically interpreted as the distance along
the magnetic field from the Sun to the Earth. Furthermore, when
combining measurements at different energies, again under the
assumption of no scattering at onset, both the injection time and
path length are estimated from a fit of the onset times and
inverse velocities to a straight line. This ``onset time versus
$1/\beta$'' method has already become a common practice
\citep{Linetal81,Reamesetal85,Kruckeretal99,KruckerL00,Tylkaetal03},
reinforced by the generally good alignment of experimental data
along a straight line in this plot.

However, the basic hypothesis of negligible scattering and motion
at zero pitch angle is hard to reconcile with the well-established
theories of particle transport. For example, using solutions of a
transport equation, \citet{KallenrodeW90} found that considerable
delays in the detected onset can arise both from interplanetary
scattering and a finite duration of the particle injection. The
onset time can also be affected by other physical processes such
as solar wind convection, and also by the technical difficulties
in measuring the onset above the pre-event particle background.

Because any major approximation of transport effects must involve
some uncertainty, we investigate the validity and systematic error
of the approximation that the first arriving particles have
undergone no scattering. We employ state-of-the-art numerical
simulations of particle transport, and analyze the resulting onset
time versus inverse velocity. We then compare the estimated start
time of injection at the Sun and path length with those actually
used in the simulation to estimate the systematic error in the
estimated values.

Validity of the inverse velocity method was also investigated by
\citet{LintunenV04}, who considered solar energetic particles of
somewhat lower energy. Where they are comparable, our results
agree quite well with those presented in their work. In general,
we find that the onset time versus $1/\beta$ method, as
investigated here for proton energies of 2~MeV to 2~GeV, has a
lower timing error than it would if applied to particles of lower
energy.

\section{Numerical experiments}

\subsection{Transport model}\label{transpsect}
We describe the propagation of protons ensuing from a solar event
by numerically solving a Fokker-Planck equation of pitch-angle
transport that includes the effects of interplanetary scattering,
adiabatic deceleration and solar wind convection
\citep{Roelof69,Ruffolo95,Nutaroetal01}. We are assuming transport
along the mean magnetic field, as expected when there is good
magnetic connection between the source and the observer. Following
\citet{NgW79}, we define the particle distribution function $F$
depending on time, $t$; pitch-angle cosine, $\mu$; distance from
the Sun along the interplanetary magnetic field, $z$; and
momentum, $p$, as:
\begin{equation}
F(t,\mu,z,p)\equiv\frac{d^3 N}{dz \, d\mu \, dp},
\end{equation}
where $N$ represents the number of particles inside a given flux
tube. The derived transport equation takes the form:
\begin{align}
\lefteqn{\frac{\pd F(t,\mu,z,p)}{\pd t} =} \nonumber \\
& \mbox{}-\frac{\pd}{\pd z}
    \mu v F(t,\mu,z,p)
            & {\rm (streaming)} \nonumber \\
& \mbox{}-\frac{\pd}{\pd z}
    \left(1-\mu^2\frac{v^2}{c^2}\right)
    v_{\rm sw}\sec\psi F(t,\mu,z,p)
            & {\rm (convection)} \nonumber \\
& \mbox{}-\frac{\pd}{\pd \mu}
    \frac{v}{2L(z)}
    \left[1+\mu\frac{v_{\rm sw}}{v}\sec\psi
        -\mu\frac{v_{\rm sw}v}{c^2}\sec\psi\right]
    (1-\mu^2) F(t,\mu,z,p)
            & {\rm (focusing)} \nonumber \\
& \mbox{}+\frac{\pd}{\pd \mu}
    v_{\rm sw}
    \left(\cos\psi\frac{d}{dr}\sec\psi\right)
    \mu(1-\mu^2) F(t,\mu,z,p)
            & {\rm (differential\ convection)} \nonumber \\
& \mbox{}+\frac{\pd}{\pd \mu}
    \, \frac{\varphi(\mu)}{2} \, \frac{\pd}{\pd\mu}
    \left(1-\mu\frac{v_{\rm sw}v}{c^2}\sec\psi\right)
    F(t,\mu,z,p)
            & {\rm (scattering)} \nonumber \\
& \mbox{}+\frac{\pd}{\pd p}
    p v_{\rm sw}
    \left[\frac{\sec\psi}{2L(z)}(1-\mu^2)
        +\cos\psi\frac{d}{dr}(\sec\psi)\mu^2\right]
    F(t,\mu,z,p).
            & {\rm (deceleration)}
\label{eqtransp}
\end{align}
Particle velocities are denoted by $v$, and the solar wind
velocity by $v_{\rm sw}$. The angle between the field line and the
radial direction is specified by the function $\psi(z)$, the
focusing length by $L(z) = -B/(dB/dz)$, and the pitch-angle
scattering coefficient by $\varphi(\mu)$.

In this work, we consider the motion of the particles along a
nominal Archimedean spiral magnetic field line \citep{Parker58}.
For this configuration
\begin{equation}
\cos \psi = \frac{R}{\sqrt{r^2+R^2}}, \quad L =
\frac{r(r^2+R^2)^{3/2}}{R(r^2+2R^2)},
\end{equation}
where $R=v_{\rm sw}/(\Omega\sin\theta_H)$, $\Omega$ is the angular
rotation rate of the Sun and $\theta_H$ the heliocentric polar
angle, and the radius $r$ can be expressed as a function of $z$.
We consider an equatorial field line ($\theta_H=\pi/2$), a typical
value for the solar wind speed ($v_{\rm sw}=400$~km~s$^{-1}$), and
a typical sidereal solar rotation period for magnetic features
($T_{\rm sid}=24.92$~days, based on a synodic period of 26.75
days; \citealt{Bai87}), so $R=0.916$~AU.

The pitch-angle scattering is parameterized as
\begin{equation}
\varphi(\mu) = A |\mu|^{q-1}(1-\mu^2). \label{pitscatt}
\end{equation}
This expression was originally derived in the context of
quasi-linear scattering theory \citep{Jokipii71,Earl73}, and here
we employ this as a convenient parameterization. The scattering
parameter $q$ is taken to be 1.5, which is consistent with values
in the range 1.3--1.7 as inferred by \citet{Bieberetal86}.

\begin{figure}
\includegraphics[width=\columnwidth]{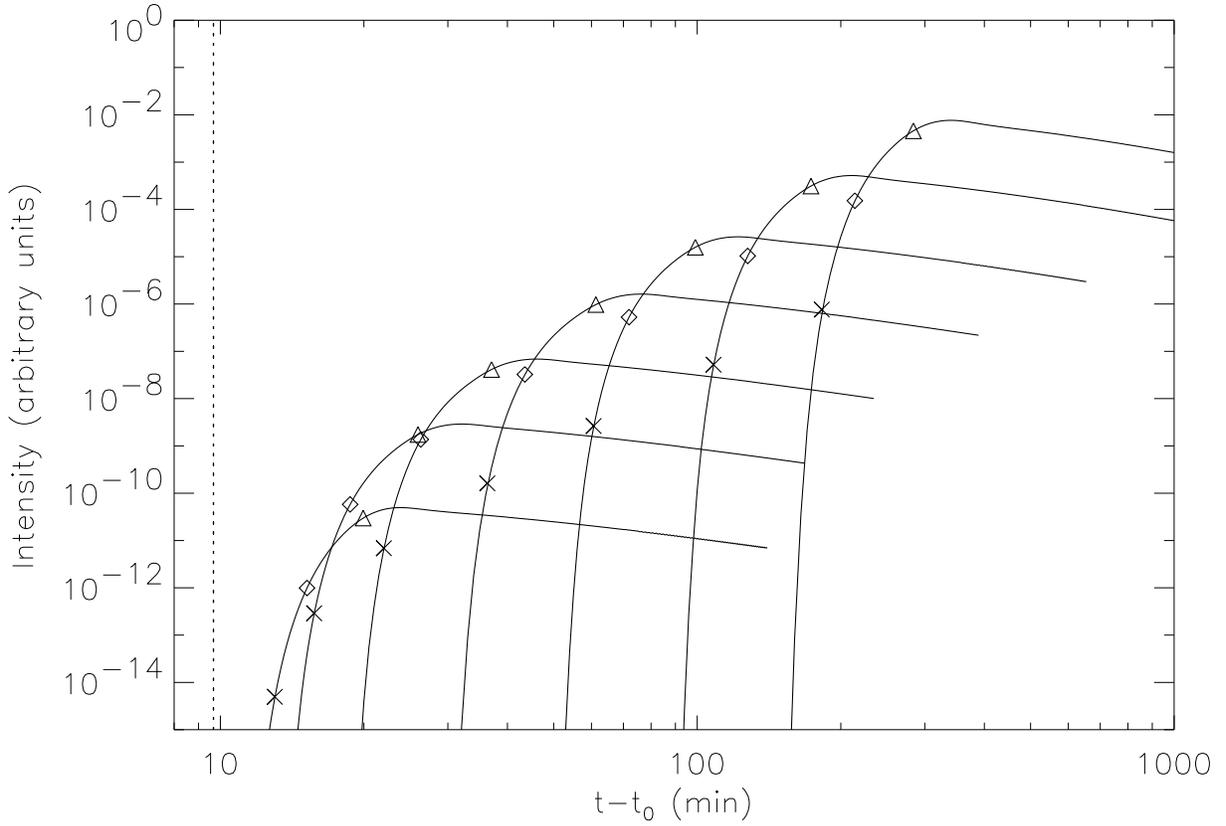}
\caption{Time profiles for the intensity at Earth orbit of protons
of different energies, resulting from an extended injection near
the Sun with a duration dependent on proton energy, and for a
constant radial mean free path $\lambda_r = 0.2$~AU. The vertical
dotted line denotes the time at which particles travelling
directly along the magnetic field line would arrive in the limit
$v \rightarrow c$. The detection times, or points where the
profiles surpass a certain threshold, are marked by crosses (low
threshold, \lowthres\ of the peak), diamonds (medium threshold,
\medthres\ of the peak), and triangles (high threshold,
\highthres\ of the peak).} \label{timeprof}
\end{figure}

The transport equation~(\ref{eqtransp}) is solved by means of a
finite-difference method \citep{Ruffolo95,Nutaroetal01}. We
consider an initial injection of particles near the Sun with zero
pitch angle ($\mu=1$). Starting from this, simulations provide us
with the intensity and anisotropy at any position along the field
line and at any time. Thus, we can construct the time profiles for
the intensity of energetic protons at the position of Earth's
orbit, for several values of the rigidity $P$. We consider seven
such values, corresponding to kinetic energies of $E=2$, 6, 20,
60, 200, 600, and 2000~MeV, that is, from non-relativistic to
relativistic proton energies. We consider only the case in which
initial intensities follow a power law in rigidity, $I \propto
P^{-5}$. An example of the resulting time profiles is depicted in
Figure~\ref{timeprof}.

\subsection{Injection profiles}\label{injprofsect}
We consider three different cases with different assumptions
regarding the rate at which protons are injected into the
interplanetary medium from the proximity of the Sun, $I(t)$.
Case~1 assumes an impulsive injection (a delta function in time)
at $t=t_0$. For Cases~2 and~3, we consider that the injection has
a time profile that peaks and decays within a certain time
interval. For simplicity, we model this by a triangular profile,
starting from $I=0$ at $t=t_0$, linearly rising to a peak at
$t=t_0+\Delta t$, and linearly returning to $I=0$ at
$t=t_0+2\Delta t$. Thus $\Delta t$ is the full width at half
maximum (FWHM) of the injection function. The value of $I$ at the
peak is chosen in such a way that the total injected intensity
remains the same as in the case of impulsive injection. In Case~2,
we assume that the profile has the same width for every value of
$P$; we take ${\rm FWHM}=12$~min. In Case~3, however, we model the
injection by considering different widths for the injection
profiles for the different rigidities. As deconvolution techniques
have shown that the injection duration tends to decrease with
increasing rigidity \citep{Ruffoloetal98,Khumlumlert01}, for this
third case we use values of the FWHM as 75, 50, 30, 20, 12, 8, and
4.8~min., for the seven energy values listed above. Note that in
all three cases, the injection is taken to start simultaneously
for every rigidity.

\subsection{Scattering mean free path}
For typical conditions of interplanetary turbulence, the diffusive
component of particle transport is characterized by some value of
the radial mean free path $\lambda_r$ on the order of 0.1 to 1~AU.
This value varies from event to event, and may also depend on
rigidity. We take into account the variability in $\lambda_r$ by
running our simulations with different assumptions for its value:
a low constant value $\lambda_r = 0.2$~AU, a high constant value
$\lambda_r = 1.0$~AU, or a value depending on rigidity, $\lambda_r
\propto P^\alpha$. We explore both positive and negative values of
the exponent, $\alpha = -1/3$, $1/3$ and 1 (in addition to the
assumptions of constant $\lambda_r$, that we can express as cases
with $\alpha=0$). In the rigidity-dependent cases, we fix the
value of $\lambda_r$ for our lowest energy, $\lambda_0$, in such a
way that the values for the other energies will roughly remain
inside the range 0.2--1.0~AU (see Table~\ref{tabivfit}). As this
is not entirely possible for $\alpha=1$, in this case we use only
five energy values instead of seven, thus not considering the
relativistic energies $E=600$ and 2000~MeV.

From quasi-linear theory, one would expect $\alpha$ to be $2-q$,
where $q$ is the power-law index of interplanetary turbulence,
also identified with the scattering parameter in
equation~(\ref{pitscatt}) \citep{Jokipii71}. For sub-GeV ions that
undergo resonant scattering with the inertial range of turbulence,
which is observed to have a Kolmogorov spectrum of turbulence with
$q=5/3$ \citep{JokipiiC68}, one might expect $\alpha=1/3$. Most
observations favor $\alpha=0$ to~$1/3$ for ions in this energy
range \citep{Palmer82,Bieberetal94}. The other values of $\alpha$
that we consider are applicable to other particle species or ions
in other energy ranges. Mean free paths inferred for solar
energetic electrons at $P<2$~MV apparently decrease with
increasing rigidity \citep{Droege00}, indicating a negative value
of $\alpha$. For ions of roughly 2 to 50~GeV, it is believed that
$q\approx 1$ and $\alpha\approx 1$ from various lines of evidence
\citep{Leerungnavaratetal03}.

\subsection{Determination of the onset time}
For each particle energy, we can define the time of detected
onset, $t_{\rm onset}$, from the simulated time profiles at Earth
orbit, as the moment when these profiles surpass a certain
threshold value. In the case of a real event, this threshold would
arise from the mean background level prior to the protons'
arrival, and its fluctuations. However, there is a large
variability in the proton background and energy spectrum before
each solar event: the background may either depend on the level of
galactic cosmic rays or be an advanced phase of a previous solar
particle event. For our purposes, we define the threshold to be a
constant fraction of the maximum intensity at each energy value.
We explore the cases in which the onset times are determined by
assuming a low threshold, corresponding to a constant fraction of
\lowthres\ of the peak, a medium threshold, at \medthres\ of the
peak, or a high threshold, at \highthres\ of the peak (see
Figure~\ref{timeprof}).

From the point of view of observations, a high threshold like
\highthres\ of the peak would be the case when the pre-event
background is very high (e.g., when a new event occurs in the
decay of a previous event), or for very small events. Such events
are not usually included in practical onset time analyses. When
conditions are favorable for observations, event onsets can be
detected at thresholds similar to our medium threshold (\medthres\
of the peak). Detection at a low threshold like \lowthres\ of the
peak would be unusual, but we include this case in order to
explore the systematic errors of the inverse velocity method in
the limit of very good observation conditions.

\section{Inverse velocity fits}\label{fitsect}
The estimation technique examined in this paper assumes that the
first detection of protons at a given energy occurs for those
protons arriving at their maximum velocity along the field line,
which would be consistent with the supposition that particles with
roughly zero pitch angle ($\mu \simeq 1$) would suffer negligible
scattering. If this holds, then $t_{\rm onset}$ as a function of
the particle velocity follows the simple relation
\begin{equation}
t_{\rm onset} = t_{\rm inj} + \frac{s}{v}, \label{oneobeta}
\end{equation}
where $t_{\rm inj}$ is the time when the injection starts and $s$
is the path length travelled by the particles. These last two
parameters can be tuned to fit the data points to the expression
in equation~(\ref{oneobeta}), which gives a straight line in the
$t_{\rm onset}$ vs.\ ${1}/{v}$ plot.

Three examples of the inverse velocity fit are shown in
Figure~\ref{onsetfit}. We see that the injection time deduced this
way (the $y$-intercept of the best-fit line) is slightly late with
respect to the actual injection start time ($t=t_0$) and that the
path length (the straight line's slope) is longer than the actual
field line length of 1.16~AU.

The results of the inverse velocity fits for all the cases studied
in this paper are summarized in Table~\ref{tabivfit} and plotted
in Figure~\ref{fitparam}. As can be seen, in most of the cases the
fits for the low threshold definition of $t_{\rm onset}$ yield an
injection time that is correct to within a few minutes and a high
value of the path length. For the higher threshold assumptions,
the trends are similar but with increased deviations from the
correct values. Our results are similar to those found by
\citet{LintunenV04} for simulated events of solar energetic protons with
energies 1--57~MeV. However, these authors found much higher
deviations (even of hours) in the estimated injection times when
considering cases with both lower proton energies (down to
130~keV) and very short mean free paths (cases that would
correspond to $\lambda_0\la 0.1$~AU).

It is worth to note that, although the range of energies that we
use throughout this work, i.e., 2--2000~MeV, is comparable to the
high-energy channels studied by \citet{LintunenV04}, i.e.,
1--57~MeV, these authors considered only one type of high-energy
dependence of $\lambda_r$ with $P$, that corresponds to our case
of $\alpha=1/3$. This leaves only one case, among the cases
studied by those authors, with which we can compare directly the
results of the present work, i.e., the one with a normalization in
the $\lambda_r$ dependence that is consistent with the one used
here (see Table~\ref{tabivfit}), which is their case no.~5,
normalized to have $\lambda_r=0.5$~AU at $P=1$~GV. Our results
indicate a deviation in the injection time of $-0.99$~min and a
path length of $1.32$~AU (when assuming a threshold of
\lowthres~of the peak), which compares well with their values,
$-1.7 \pm 0.4$~min and $1.33$~AU (for a threshold of 0.1\%~of the
peak).

We now study how the different simulation parameters affect the
estimated parameters $t_{\rm inj}$ and $s$. For example, regarding
the injection profile, we observe that the trend is similar for
each mean free path assumption: if the injection is extended and
of constant width (${\rm FWHM}=12$~min), the derived injection
time, $t_{\rm inj}$, is always later than in the impulsive
injection case, and the path length, $s$, changes slightly. When
the width of the injection profile varies, with broader injections
for lower rigidities, $t_{\rm inj}$ is usually slightly later but
not far from the impulsive injection case. On the other hand,
derived $s$ values are systematically longer. These differences
for different injection assumptions also increase for the higher
threshold assumptions. Our results are consistent with those of
\citet{LintunenV04}, who studied only the effects of an extended
injection with constant width, and who also found a delay in
derived injection times and slightly longer path lengths. The
results are qualitatively the same, although these authors used a
much longer injection (${\rm FWHM}\ge 3.7$~h) and a ``top-hat''
profile.

As previously mentioned, the injection time, $t_{\rm inj}$, is
usually not far from the time of injection used in the
simulations, $t_0$, but we observe that simulation results for the
case in which the scattering mean free path decreases with
rigidity (negative $\alpha$) tend to give later injection times
while the cases for $\lambda$ increasing with rigidity (positive
$\alpha$) tend to give earlier injection times. This is
particularly extreme for $\alpha=1$, a case of strong dependence.

The estimated path length, $s$, for an impulsive injection seems
to depend mainly on the value of the mean free path at the lowest
energy. This is because the point corresponding to the onset of
the slowest particles will have a higher relative weight in the
$t_{\rm onset}$ vs.\ $1/v$ fit. This is also apparent in the
results of \citet[their Table~2 and Figure~3]{LintunenV04}, though
not explicitly discussed in that work. In cases where a different
low-energy dependence of $\lambda_r$ with $P$ resulted in a
similar value of $\lambda_r$ for the lowest energy value (e.g.,
their cases 6 and~13), the resulting value of $s$ was similar.

An interesting result is that while many of the simulations yield
estimated $t_{\rm inj}$ and $s$ values with substantial errors,
all the simulations are nevertheless well fit by straight lines in
$t_{\rm onset}$ versus $1/v$.  The worst fit of all is included in
Figure~\ref{onsetfit} (indicated by triangles), and to the eye
even this appears quite good.  (Actual observations would involve
some fluctuations, which may even dominate the deviations shown
here.) Yet according to Figure~\ref{fitparam}, some of these
apparently good fits yield path lengths over 2~times too long or
injection time errors of over 10~minutes. Thus the goodness of fit
alone does not validate the results or assumptions of the inverse
velocity fits.

To quantitatively compare the various fits, it is interesting to
examine the unweighted $\chi^2$ function,
\begin{equation}
\chi^2=\sum_{i=1}^N{
    \left[ t_{{\rm onset,}i} - \left( t_{\rm inj} + \frac{s}{v_i} \right) \right] ^2
    }.
\end{equation}
We can regard this not only as an estimation of the goodness of
the fit but also as a measure of the departure of our
``experimental data'' from the linear relation. In
Table~\ref{tabivfit} we show the value of $\chi^2$ divided by the
number of degrees of freedom, $N-2$, with $N$ as the number of
data points. We find that the worst fits are mostly among the
cases of strong dependence of $\lambda_r$ with rigidity, that is,
for $\alpha=1$ and $-1/3$, although there are exceptions. On the
other hand, one of the best fits, which is also the point
appearing closest to the correct values in Figure~\ref{fitparam},
is that corresponding to an impulsive injection and a high
constant mean free path, $\lambda_r=1.0$~AU. This is actually the
set of parameters closest to those that would fulfill
Eq.~(\ref{oneobeta}) exactly (in the absence of solar wind
convection) for any threshold, that is, an impulsive injection
with a mean free path $\lambda_r\rightarrow\infty$.

Although all of the linear fits are good, it is interesting to
note that a very good fit, or even the best fit (depending on the
threshold assumption), corresponds to the case for which
$\lambda_r$ increases with rigidity as $P^\alpha$ with
$\alpha=1/3$. This very good alignment arises from offsetting
effects of the variable $\lambda_r$ and the solar wind convection,
the former favoring the transport of faster particles and the
latter having a greater relative importance for slower particles.
[Note also that $\alpha=1/3$ would arise from quasi-linear theory
with a Kolmogorov spectrum of magnetic fluctuations
\citep{Jokipii71} and is also supported observationally
\citep[e.g.,][]{ValdesGaliciaetal95} and theoretically
\citep[e.g.,][]{Bieberetal94}.] However, even in this case, the
fit parameters differ considerably from the correct values,
especially the path length of 1.3--1.7~AU, depending on the
threshold. This means that for typical values of the solar wind
speed, the interplanetary turbulence effects on the transport of
particles of different energies can produce onset times near Earth
that align extremely well in the inverse velocity plot, but whose
fit leads to a value of the path length 20--50\% longer than the
actual length of the field line.

This may explain some of the results of \citet{Tylkaetal03}, who
found long path lengths in some events analyzed using the onset
time versus $1/\beta$ method. They included measurements of
protons, ions, and electrons for $1/\beta$ values of $\simeq
1$--15, which for protons corresponds to energies from $\simeq
2$~MeV to several~GeV. Using this method they found, e.g., a path
length of $1.36 \pm 0.02$~AU for the impulsive event of May~1,
2001, and a path length of $1.67 \pm 0.02$~AU for the ground level
event of April~15, 2001. This last event (known as the Easter 2001
event) was also studied by \citet{Bieberetal04}, who found a good
fit to neutron monitor data by using a detailed treatment of
interplanetary transport (as described in \S\ref{transpsect}) and
a magnetic field line length of approximately 1.2~AU, as expected
given the value of $v_{\rm sw}$ during that event. We have
performed a similar analysis under the supposition of a path
length of 1.7~AU, but in this case the fit to the data is quite
poor, with a value of $\chi^2$ per degree of freedom that is
roughly 9~times larger than that found by \citet{Bieberetal04} for
1.2~AU. Note that the injection start times derived by
\citet{Tylkaetal03} and \citet{Bieberetal04} agree to within
3~minutes. The small difference in injection time and major
difference in path length is entirely consistent with the results
of the present work.

\section{Implications for solar electrons}

The study of solar energetic electrons is of special interest in
understanding particle emission and transport in solar events. The
propagation of solar electrons can be tracked by Type III radio
emission \citep{ReinerS88,ReinerS89} and has been shown to be
closely related to the propagation of ions \citep{CaneE03}. In
addition, electron onset times are generally detected with better
accuracy than for ions, and are also used to infer injection times
through the inverse velocity method
\citep[e.g.,][]{Linetal81,Kruckeretal99}.

Although in the present work we focus on the transport of solar
protons, the question arises at this point about whether the
results of \S\ref{fitsect} can be applied to solar electrons as
well. Our answer is yes. The transport of solar energetic
particles along the magnetic field line that connects the
acceleration site to the observer does not depend on the
particles' mass or the sign of their electric charge. It does
depend on their velocities, on the interplanetary scattering
conditions, and to a lesser extent, on the spectral index of
particle emission, see equation~(\ref{eqtransp}). Therefore, the
results presented above are applicable, under the same assumptions
regarding scattering mean free paths $\lambda_r$ and injection
profiles, to solar electrons of similar velocities as the seven
values used for protons, which correspond, roughly speaking, to
electron energies of $E=1$, 3, 10, 30, 100, 300, and 1000~keV.
This range of energies is also compatible with most solar
energetic electron observations.

From these considerations, we also conclude that the onset
analysis of electrons and ions for the same event may lead to
differences in the fit parameters that would arise from
differences in $\lambda_r$, in the injection duration, or in the
threshold for detecting the onset. It has been shown
\citep[especially Figure~3]{Bieberetal94} that the electron and
proton mean free paths are similar in the same event for a range
of mean free paths over two orders of magnitude. Even if we
consider that the value of $\lambda_r$ is exactly the same, the
inverse velocity method will give different results for protons
and electrons if their injection durations are different. If, for
example, the injection is impulsive for electrons but extended
with constant width for protons, and both injections start
simultaneously, the fits will estimate a later $t_{\rm inj}$ for
protons than for electrons, and approximately the same $s$. Under
the same conditions except that the injection duration for protons
depends on rigidity as proposed in \S\ref{injprofsect}, the
resulting $t_{\rm inj}$ will be similar in both fits but $s$ will
be longer for protons. Further deviations may arise from
differences in the rigidity dependence of $\lambda_r$. The results
of \citet{Droege00} suggest that the mean free path varies
smoothly with rigidity from electrons at low rigidity to protons
at higher rigidity for the same event, but changes its dependence
with rigidity, with negative $\alpha$ for electrons and positive
$\alpha$ for protons. According to our results, this effect would
add an extra difference in $s$ between protons and electrons, with
longer $s$ for protons. We refer the reader to
Table~\ref{tabivfit} for other possible combinations. In summary,
the inverse velocity method may introduce artificial differences
in estimated injection times and especially path lengths when used
with electrons and protons for the same event
\citep[e.g.,][]{KruckerL00}.

\section{Discussion and Conclusions}

The results of the pitch-angle transport simulations are not
compatible with the assumption of no scattering for the first
detected particles. We find that the onset time for each energy is
always influenced by interplanetary scattering, the duration of
injection, and solar wind convection, each of which may have a
different relative importance for different energies.  The onset
delays found for single energy values are consistent with
\citet{KallenrodeW90}, who used a focused transport equation but
did not include any effects of the solar wind speed.

By combining the onset times at different energies in the
inverse-velocity plot, a simple fit leads to an estimation of the
start time of the injection, $t_{\rm inj}$, and the path length,
$s$, with results comparable to those of \citet{LintunenV04}. We
find that often a good linear fit is obtained fortuitously, even
when the fit parameters deviate substantially from the true
values. Thus the goodness of the fit should not be taken as an
indication that the estimated values or underlying assumptions are
correct.

While \citet{LintunenV04} found some cases with a major error in
the timing estimation (even of hours), this was never the case
when they considered only proton energies greater than 1~MeV. We do not
find deviations in the timing estimation larger than several
minutes in our results for protons of 2~MeV to 2~GeV, or electrons of 1~keV to 1~MeV. We note that
most practical applications of this estimation technique
\citep[e.g.,][]{Linetal81,Reamesetal85,Kruckeretal99,Tylkaetal03}
have considered values of $1/\beta$ up to only 10 or 20,
corresponding to proton energies larger than 1~MeV, or electron energies larger than 500~eV. In such cases,
the onset time versus $1/\beta$ method is unlikely to incur timing
errors greater than several minutes.

It should be noted that fits to intensity and anisotropy data of solar energetic particles
using detailed transport modeling can yield a complete injection
function, not only the start time of injection. Such information
can be derived for each energy, as well as the mean free path and
magnetic configuration \citep{Ruffoloetal98,Bieberetal02}. On the
other hand, although the use of inverse-velocity fits is based on
a highly simplified assumption for interplanetary transport, it
often results in roughly correct start times of injection, with
typical errors of several minutes, as long as the interplanetary
scattering is not excessively high.

\begin{figure}
\includegraphics[width=\columnwidth]{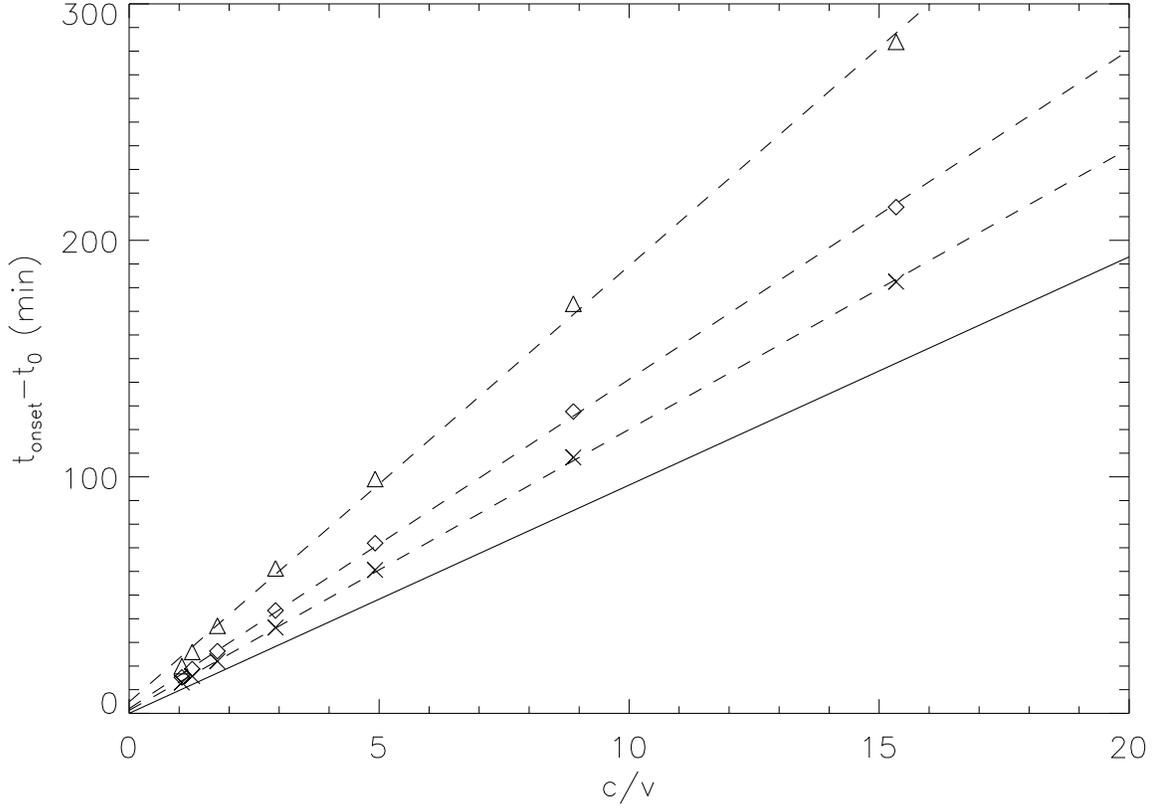}
\caption{Inverse velocity fit to the onset times of protons of
different energies, deduced for the case shown in
Figure~\ref{timeprof}, showing all three threshold assumptions.
The linear fits (dashed lines) correspond to $1.25{\rm
~min}+1.43{\rm ~AU}/v$ for the low threshold (crosses, \lowthres\
of the peak), $1.99{\rm ~min}+1.67{\rm ~AU}/v$ for the medium
threshold (diamonds,  \medthres\ of the peak), and $4.82{\rm
~min}+2.22{\rm ~AU}/v$ for the high threshold (triangles,
\highthres\ of the peak). The solid line is $0.00{\rm
~min}+1.16{\rm ~AU}/v$, i.e., where points would line up if
particles could travel directly along the magnetic field without
any scattering.} \label{onsetfit}
\end{figure}

\begin{figure}
\includegraphics[width=\textwidth]{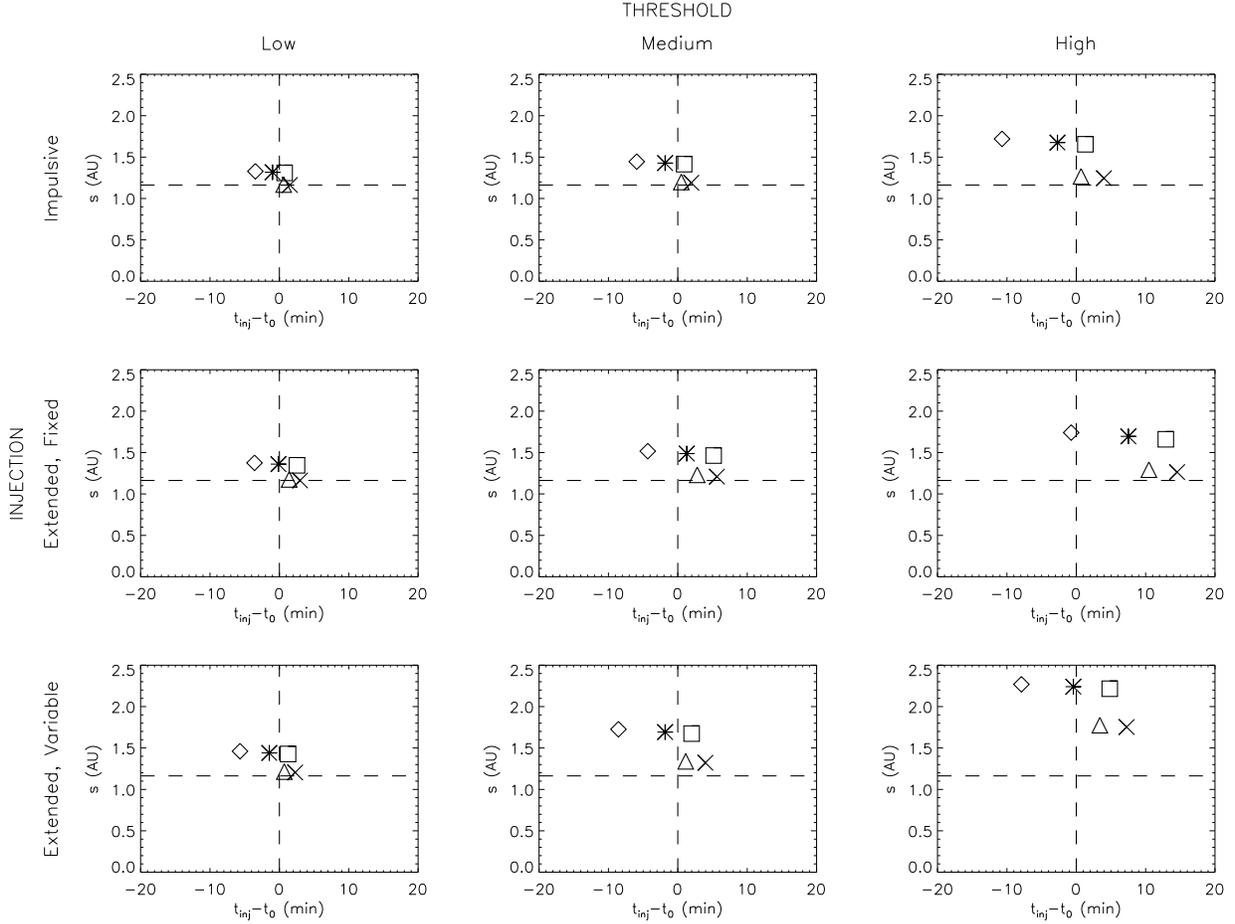}
\caption{Times of injection vs.\ path lengths estimated from the
inverse velocity fits in the different cases studied. Left panels:
results of the low threshold assumption (\lowthres\ of the peak).
Center panels: results of the medium threshold assumption
(\medthres\ of the peak). Right panels: results of the high
threshold assumption (\highthres\ of the peak). Top panels:
impulsive injection. Middle panels: extended injection with
constant absolute width (12~min). Bottom panels: extended
injection with width depending on particle rigidity (see text).
Different symbols denote different mean free path assumptions:
crosses: $\lambda_0=1.0$~AU, $\alpha=-1/3$; triangles:
$\lambda_0=1.0$~AU, $\alpha=0$; squares: $\lambda_0=0.2$~AU,
$\alpha=0$; stars: $\lambda_0=0.2$~AU, $\alpha=1/3$; diamonds:
$\lambda_0=0.2$~AU, $\alpha=1$. The intersection of dashed lines
indicates the actual values of the start time of injection and
path length used in performing all the simulations.}
\label{fitparam}
\end{figure}

In contrast, the path lengths obtained from the linear fits are
frequently very different from the actual path length along the
local interplanetary magnetic field line, and always larger. This
clearly indicates that the first arriving particles cannot be
viewed as moving directly along the interplanetary magnetic field,
with $v_z=v$ and zero pitch angle.  Empirically, a better
assumption would be that the first arriving particles travel with
an energy-independent value of the pitch angle cosine,
$\mu=v_z/v$, that is less than unity. Indeed, this is a feature of
the coherent pulse concept of focused transport theory
\citep{Earl76a,Earl76b} for a scattering mean free path that
depends only weakly on energy.  Figure~\ref{onsetfit} would seem
to imply that higher thresholds (relative to the peak) represent
the arrival of particles with decreasing $\mu$.  For a path length
$z$, the ``onset'' particles empirically arrive roughly at time
\begin{equation}
t_{\rm onset} = t_{\rm inj} + \frac{z}{v_z} = t_{\rm inj} +
\frac{s}{v},
\end{equation}
where $s=z/\mu$ for an unknown value of $\mu$ that is apparently
roughly constant with energy.  The coherent pulse concept can help
explain why the inverse velocity fits provide estimates of $s$
that do not represent the actual path length, but rather the path
length magnified by some factor ($1/\mu$), and estimates of the
start time of injection that are accurate to the order of several
minutes.

\acknowledgments This work is based on a presentation to the
ACE/RHESSI/WIND Workshop in Taos, NM on October 8, 2003. The
authors acknowledge useful discussions at the 2003 SHINE Workshop
and with A.~Tylka. We thank P.~Wongpan for his assistance. This
work was partially supported by the Thailand Research Fund, the
Rachadapisek Sompoj Fund of Chulalongkorn University, and the US
National Science Foundation (award ATM--0000315).

\bibliographystyle{apj}
\bibliography{SEP}

\clearpage

\begin{deluxetable}{cccccrcr}
 \tablecaption{Results of the inverse velocity fits\label{tabivfit}}
 \tablehead{
  \colhead{$\lambda_0$\tablenotemark{a}} & \colhead{$\alpha$} &
  \colhead{Injection} & \colhead{Duration} & \colhead{Threshold} &
  \colhead{$t_{\rm inj}-t_0$} & \colhead{$s$} &
  \colhead{$\chi^2/(N-2)$}\\
  \colhead{(AU)} & & & & \colhead{(\% of peak)} & \colhead{(min)} & \colhead{(AU)} &
  \colhead{(min$^2$)}
  }
\startdata
1.0&  $-$1/3\phs& Impulsive&       --& 0.01& $  1.52$\phn\phn&  1.16&   0.39\phn\phn\\
1.0&  $-$1/3\phs&  Extended& Constant& 0.01& $  2.98$\phn\phn&  1.16&   0.96\phn\phn\\
1.0&  $-$1/3\phs&  Extended& Variable& 0.01& $  2.28$\phn\phn&  1.20&   0.58\phn\phn\\
1.0&   \phs0\phs& Impulsive&       --& 0.01& $  0.62$\phn\phn&  1.17&   0.16\phn\phn\\
1.0&   \phs0\phs&  Extended& Constant& 0.01& $  1.41$\phn\phn&  1.18&   1.07\phn\phn\\
1.0&   \phs0\phs&  Extended& Variable& 0.01& $  0.70$\phn\phn&  1.21&   0.26\phn\phn\\
0.2&   \phs0\phs& Impulsive&       --& 0.01& $  0.76$\phn\phn&  1.31&   0.27\phn\phn\\
0.2&   \phs0\phs&  Extended& Constant& 0.01& $  2.54$\phn\phn&  1.35&   1.35\phn\phn\\
0.2&   \phs0\phs&  Extended& Variable& 0.01& $  1.25$\phn\phn&  1.43&   0.94\phn\phn\\
0.2& \phs1/3\phs& Impulsive&       --& 0.01& $ -0.99$\phn\phn&  1.32&   0.04\phn\phn\\
0.2& \phs1/3\phs&  Extended& Constant& 0.01& $ -0.13$\phn\phn&  1.36&   0.19\phn\phn\\
0.2& \phs1/3\phs&  Extended& Variable& 0.01& $ -1.45$\phn\phn&  1.44&   0.03\phn\phn\\
0.2&   \phs1\phs& Impulsive&       --& 0.01& $ -3.46$\phn\phn&  1.33&   2.00\phn\phn\\
0.2&   \phs1\phs&  Extended& Constant& 0.01& $ -3.58$\phn\phn&  1.38&   1.31\phn\phn\\
0.2&   \phs1\phs&  Extended& Variable& 0.01& $ -5.68$\phn\phn&  1.46&   3.87\phn\phn\\
1.0&  $-$1/3\phs& Impulsive&       --&  2.0& $  1.94$\phn\phn&  1.19&   0.21\phn\phn\\
1.0&  $-$1/3\phs&  Extended& Constant&  2.0& $  5.64$\phn\phn&  1.21&   1.13\phn\phn\\
1.0&  $-$1/3\phs&  Extended& Variable&  2.0& $  3.97$\phn\phn&  1.32&   2.03\phn\phn\\
1.0&   \phs0\phs& Impulsive&       --&  2.0& $  0.50$\phn\phn&  1.20&   0.09\phn\phn\\
1.0&   \phs0\phs&  Extended& Constant&  2.0& $  2.80$\phn\phn&  1.23&   0.91\phn\phn\\
1.0&   \phs0\phs&  Extended& Variable&  2.0& $  1.14$\phn\phn&  1.34&   0.75\phn\phn\\
0.2&   \phs0\phs& Impulsive&       --&  2.0& $  0.91$\phn\phn&  1.41&   0.39\phn\phn\\
0.2&   \phs0\phs&  Extended& Constant&  2.0& $  5.16$\phn\phn&  1.46&   2.32\phn\phn\\
0.2&   \phs0\phs&  Extended& Variable&  2.0& $  1.99$\phn\phn&  1.67&   2.32\phn\phn\\
0.2& \phs1/3\phs& Impulsive&       --&  2.0& $ -1.82$\phn\phn&  1.43&   0.15\phn\phn\\
0.2& \phs1/3\phs&  Extended& Constant&  2.0& $  1.30$\phn\phn&  1.49&   0.66\phn\phn\\
0.2& \phs1/3\phs&  Extended& Variable&  2.0& $ -1.82$\phn\phn&  1.69&   0.12\phn\phn\\
0.2&   \phs1\phs& Impulsive&       --&  2.0& $ -5.92$\phn\phn&  1.45&   6.84\phn\phn\\
0.2&   \phs1\phs&  Extended& Constant&  2.0& $ -4.34$\phn\phn&  1.52&   2.00\phn\phn\\
0.2&   \phs1\phs&  Extended& Variable&  2.0& $ -8.56$\phn\phn&  1.73&   6.05\phn\phn\\
1.0&  $-$1/3\phs& Impulsive&       --& 60.0& $  3.99$\phn\phn&  1.25&   0.93\phn\phn\\
1.0&  $-$1/3\phs&  Extended& Constant& 60.0& $ 14.54$\phn\phn&  1.26&   2.41\phn\phn\\
1.0&  $-$1/3\phs&  Extended& Variable& 60.0& $  7.26$\phn\phn&  1.75&  13.81\phn\phn\\
1.0&   \phs0\phs& Impulsive&       --& 60.0& $  0.69$\phn\phn&  1.27&   0.22\phn\phn\\
1.0&   \phs0\phs&  Extended& Constant& 60.0& $ 10.47$\phn\phn&  1.29&   1.42\phn\phn\\
1.0&   \phs0\phs&  Extended& Variable& 60.0& $  3.39$\phn\phn&  1.78&   8.06\phn\phn\\
0.2&   \phs0\phs& Impulsive&       --& 60.0& $  1.31$\phn\phn&  1.66&   0.79\phn\phn\\
0.2&   \phs0\phs&  Extended& Constant& 60.0& $ 12.90$\phn\phn&  1.66&   1.05\phn\phn\\
0.2&   \phs0\phs&  Extended& Variable& 60.0& $  4.82$\phn\phn&  2.22&  15.35\phn\phn\\
0.2& \phs1/3\phs& Impulsive&       --& 60.0& $ -2.72$\phn\phn&  1.68&   0.07\phn\phn\\
0.2& \phs1/3\phs&  Extended& Constant& 60.0& $  7.51$\phn\phn&  1.70&   0.26\phn\phn\\
0.2& \phs1/3\phs&  Extended& Variable& 60.0& $ -0.42$\phn\phn&  2.24&   4.46\phn\phn\\
0.2&   \phs1\phs& Impulsive&       --& 60.0& $-10.70$\phn\phn&  1.72&  11.45\phn\phn\\
0.2&   \phs1\phs&  Extended& Constant& 60.0& $ -0.74$\phn\phn&  1.74&   5.42\phn\phn\\
0.2&   \phs1\phs&  Extended& Variable& 60.0& $ -7.92$\phn\phn&  2.27&   2.74\phn\phn\\

\enddata
\tablenotetext{a}{Radial mean free path at a proton kinetic energy of 2~MeV (or a electron kinetic energy of $\approx 1$~keV, see text).}
\end{deluxetable}

\clearpage

\end{document}